# Can animal manure be used to increase soil organic carbon stocks in the Mediterranean as a mitigation climate change strategy?

Andreas Kamilaris[1,2], Immaculada Funes Mesa [3], Robert Savé[3], Felicidad De Herralde[3] and Francesc X. Prenafeta-Boldú[3]

[1] Research Centre on Interactive Media, Smart Systems and Emerging Technologies (RISE), Nicosia, Cyprus
[2] Department of Computer Science, University of Twente, The Netherlands
[3] Institute of Agrifood Research and Technology, Barcelona, Spain

**Abstract.** Soil organic carbon (SOC) plays an important role on improving soil conditions and soil functions. Increasing land use changes have induced an important decline of SOC content at global scale. Increasing SOC in agricultural soils has been proposed as a strategy to mitigate climate change. Animal manure has the characteristic of enriching SOC, when applied to crop fields, while, in parallel, it could constitute a natural fertilizer for the crops. In this paper, a simulation is performed using the area of Catalonia, Spain as a case study for the characteristic low SOC in the Mediterranean, to examine whether animal manure can improve substantially the SOC of agricultural fields, when applied as organic fertilizers. Our results show that the policy goals of the 4x1000 strategy can be achieved only partially by using manure transported to the fields. This implies that the proposed approach needs to be combined with other strategies.

**Keywords:** Soil Organic Carbon Stock, Animal Manure, Climate Change.

## 1     Introduction

Soil organic carbon (SOC) plays a crucial role on physical, chemical and biological soil conditions and consequently on soil functions such as sustaining and enhancing crop yields due to the increase in water and nutrients storage and availability for plants [1]. However, an important decline of SOC content has been observed at the global scale because of land use changes, like deforestation in favour of cultivation and intensive agricultural practices [2]. SOC depletion is basically promoted by changes in soil temperature and texture, moisture regimes, soil disturbance and erosion [1].
Estimates indicate that SOC is being lost globally at an annual rate equivalent to 10–20% of the total global carbon dioxide emissions. Most soil landscapes in the southern and eastern parts of the Mediterranean basin, independently from the cropping systems, are subject to SOC losses which are mainly attributable to relatively high temperatures and soil erosion. In the northern part of the basin, the issue of low SOC stocks is of particular concern in perennial systems such as orchards and vineyards [3], which play an important socioeconomic role in southern Europe.

Recent data showed that bare soils, vineyards and orchards in Europe are prone to erosion (10-20 tonnes ha$_{-1}$ yr$_{-1}$), while cropland and fallow show smaller soil losses (6.5 and 5.8 tonnes ha$_{-1}$ yr$_{-1}$) largely because the latter occupy land with little or no slope. Grasslands and the associated livestock rearing are of limited extent in Mediterranean regions, so the accumulation of SOC associated with such land uses is severely restricted. Overgrazing is a potential threat though. In addition, wildfires, which are rather common in the Mediterranean, can also have a negative impact on SOC, but they normally affect forests and rangelands and are thus of limited concern in cultivated agro-ecosystems. In the particular case of Spain, there was a continued decline in SOC during the 20th century (cropland SOC levels in 2008 were 17% below their 1933 peak) and these SOC trends were driven by historical changes in land uses, management practices and climate [4]. In this sense, the FOOD chapter of the first MedECC report focused on the loss of soil organic carbon (SOC) [5]. Fortunately, certain soil management practices may significantly influence the capacity of the soil to sequester SOC in agricultural soils [1]. Several broad strategies have been suggested to increase or maintain soil fertility (water storage, nutrients source, biodiversity preservation, etc.) since the very start of agriculture, although the recent intensification has been distorting soil fertility. The main strategies to increase SOC contents and fertility in Mediterranean soils include reduced soil tillage, crop rotations, cover crops and the introduction of new crop varieties or cultures, among others. In addition, synergies and trade-offs between adaptation and mitigation strategies must be considered for an enhanced resilience of agro-systems, particularly in regions highly impacted by climate change such as the Mediterranean [6], [7].

In this context, the international initiative "*4 per 1000 Soils for Food Security and Climate*" was launched by France on 1st December 2015 at the 21st Conference of the Parties to the United Nations Framework Convention on Climate Change (COP21) and validated on November 16 in Marrakech (COP22). The "4 per 1000" (4x1000) initiative consists of federating all voluntary stakeholders of the public and private sectors (national governments, local and regional governments, companies, trade organisations, NGOs, research facilities, etc.) under the framework of the Lima-Paris Action Plan (LPAP). The aim of the initiative is to demonstrate that agriculture, and in particular agricultural soils, can play a crucial role where food security and climate change are concerned by increasing soil organic matter stocks by 4 per 1000 (or 0.4 %) per year as a compensation for the global emissions of greenhouse gases by anthropogenic sources [8]. This strategy has been included within the agronomic practices focused on the mitigation and adaptation of agriculture to climate change in Catalonia (Northeast of Spain). Supported by solid scientific documentation, this initiative invites all participants and stakeholders to state and implement specific practices aimed at enhancing soil carbon storage and the type of practices to achieve this (e.g. agroecology, agroforestry, conservation agriculture, landscape management, etc.). The ambition of the initiative is to encourage stakeholders to transition towards a productive, highly resilient agriculture, based on the appropriate management of lands and soils, creating jobs and incomes hence ensuring sustainable development.

The present study assesses the application of organic matter in agricultural soils as an option of carbon sequestration strategy at the regional scale, considering as well the problems associated to organic matter characteristics, transportation and application in a global carbon and economic footprint have also been considered in this assessment. This will allow to assess the feasibility of reaching the 4x1000 target by means of animal manure used as agricultural fertilizer.

## 2      Related Work

Surplus manure from livestock has been used in the past as fertilizer in crop fields. An approach for the transportation of manure beyond individual farms for nutrient utilization was proposed in [9], focusing on animal manure distribution in Michigan.
Teira-Esmatges et al. [10] proposed a methodology to distribute manure at a regional and municipal scale in an agronomically correct way, i.e. by balancing manure application based on territorial crop needs, as well as on predictions of future needs and availability considering changes in land use. ValorE [11] is a GIS-based decision support system for livestock manure management, with a small case study performed at a municipality level in the Lombardy region, northern Italy, indicating the feasibility of manure transfer. Other researchers proposed approaches to select sites for safe application of animal manure as fertilizer to agricultural land [12-13]. Site suitability maps have been created using a GIS-based model in the Netherlands and in Queensland, Australia respectively.
Several studies reported agricultural management practices that are found to be beneficial to sequester carbon from various regions in the world (see the literature review and listed sequestration rates in [8]. Addition of organic amendments to the soil is one of the more reported management practices to increase SOC, along with other agricultural practices [14-18] such us reduced tillage [19-20], crop residue incorporation [20-22], cover crops [23-24], crop rotation [20], [25] or land use changes [26-28].
Moreover, organic farming [4], [29-31] and conservation agriculture [32] are reported to effectively promote carbon sequestration when compared to conventional agriculture. Most of these recommended practices are supposed to increase SOC by themselves but some studies have highlighted the potential of combining practices to improve sequestration [16-18]. Moreover, there are studies on upscaling the effects of SOC on climate change, land use changes or the application of certain agricultural practices [15], [26], [33]. However, compost or organic amendments large scale estimates for the carbon sequestration potential of compost and organic amendments in agricultural soils are still scarce [18].

## 3      Methodology

The purpose of this section is to describe how the problem was modelled using the area of Catalonia as a case study.

### 3.1 Problem description

Catalonia is one of the European regions with the highest livestock density, according to the agricultural statistics for 2016, provided by the Ministry of Agriculture, Government of Catalonia. The animal census reported numbers of around 9M pigs, 0.7M cattle and 75M poultry in a geographical area of 32,108 square kilometers, hosted in 25K livestock farms. The overall goal is to solve the problem of how to find the optimal and economically viable way to distribute animal manure in order to fulfil agricultural fertilization needs in one hand and to increase carbon stock of these fields on the other hand. To simplify the problem, the geographical area of Catalonia has been divided into a two-dimensional grid, as shown in Figure 1 (left). In this way, the distances between livestock farms (i.e. original grid cell) and crop fields (e.g. destination grid cell) are easier to compute, considering straight-line grid cell Manhattan distance as the metric to use and not actual real distance through the existing transportation network. Each crop field and livestock farm has been assigned to the grid cell where the farm is physically located.

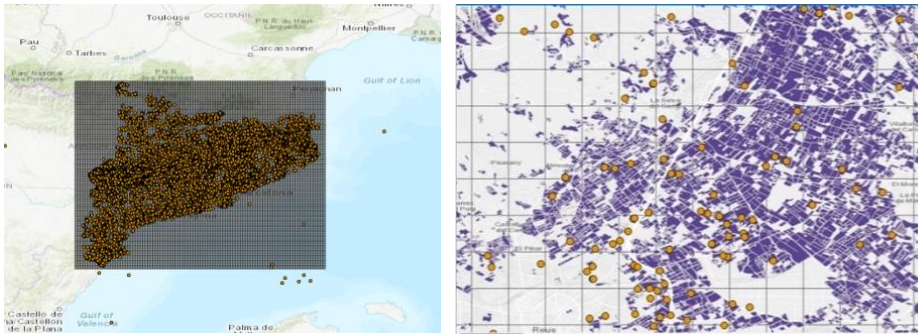

**Fig. 1.** Division of the territory of Catalonia in cells of 1 square kilometer. each (left). Demonstration of livestock farms and crop fields at grid cells in a dense agricultural area of the region (right). This is a zoom of the map shown on the left. Livestock farms are shown as brown circles, and crop fields as blue polygons.

### 3.2 Data collection and pre-processing

Details about livestock farms (i.e. animal types and census, location etc.) have been provided by the Department of Agriculture of the Government of Catalonia for the year 2016, after signing a confidentiality agreement. Details about crop fields (i.e. crop type, hectares, irrigation method, location etc.) have been downloaded from the website of the Department [34], for the year 2015. More than 20K crop fields have been recorded. For every livestock farm, the yearly amount of manure produced and its equivalent in nitrogen as fertilizer have been calculated, depending on the type and number of animals on the farm, based on the IPCC guidelines (TIER1) [35]. Similarly, for every crop field, the yearly needs in nitrogen have been computed, depending on the crop type and total hectares of land, according to the Nitrate Directive [36]. The estimated total fertilizer needs of crop fields (i.e. 88K tons of nitrogen) were lower than the availability of nitrogen from animal manure (i.e. 116K tons of nitrogen) [37-38]. This means that the produced amount of manure/nitrogen from livestock agriculture has the potential to completely satisfy the total needs of crop farms. Further, we needed to know the existing soil organic content (SOC) in soils around Catalonia and to associate this with the existing crop fields. This information was retrieved from the SOC stock baseline map for Catalonia, published in [39] and downloaded from the Institut Cartogràfic i Geològic de Catalunya (ICGC) [40]. The correlation with crop fields was performed in ESRI ArcGIS. Finally, it was necessary to estimate the carbon content by applying animal manure to the Catalonian soils. Table 1 shows the relevant parameters used to estimate the kilograms of carbon contained in animal manure in terms of volume ($m_3$) or weight (kg). The values are taken from [49-51].

**Table 1**. Equivalence table of relevant parameters for each animal fertilizer type to estimate carbon contained in animal manure.

|  | Density$_1$ (ton/m3) | Relative unit | Dry matter$_2$ (kg) | Organic matter$_2$ (kg) | Total organic Carbon (kg C/m$_3$) | | kg C/kg fertilizer |
|---|---|---|---|---|---|---|---|
| Fertilizer type |  |  |  |  | Range | Average value | Average Value |
| Pig slurry (fattening) | 1.050 | Per 1 m$_3$ | 70 | 40 – 50 | 20.86 - 26.08 | 23.47 | 0.0224 |
| Pig slurry (maternity) | 1.024 | Per 1 m$_3$ | 50 | 20 - 30 | 10.43 - 15.65 | 13.04 | 0.0127 |
| Manure (bovine) | 0.750 | Per 1 ton | 210 | 120 - 150 | 62.60 – 78,25 | 70.42 | 0.07042 |
| Manure (poultry) | 0.850 | Per 1 ton | 600 | 370 - 420 | 193.01 – 219.09 | 206.05 | 0.20605 |
| Compost (bovine manure)$_3$ | 0.720 | Per 1 ton | 508 | 274.8 | - | 143.35 | 0.14335 |
| Compost (poultry manure)$_3$ | 0.720 | Per 1 ton | 754 | 418.4 | - | 218.26 | 0.21826 |

### 3.3 Modelling

The total area of Catalonia has been divided into 74,970 grid cells, each representing a 1x1 square kilometer of physical land. Every cell has a unique ID and (x,y) coordinates, ranging between [1,315] for the x coordinate and [1,238] for the y coordinate.

For each grid cell, we are aware of the crop and livestock farms located inside that cell, the manure/nitrogen production (i.e. from the livestock farms) and the needs in nitrogen (i.e. of the crop fields). Moreover, some crop farms in Catalonia fall within *nitrate vulnerable zones*. Inside these zones, only a maximum of *170 kg/ha* (i.e. kilos/hectare) of nitrogen from manure can be applied. We assumed that each transport vehicle (i.e. truck) used for the transfer of manure has a limited capacity of 20 cubic meters to transfer manure/nitrogen. The allowed periods of the year when fertilizer can be applied on the land (depending whether the crop falls within a vulnerable area or not) based on the Directive 153/2019 of the Government of Catalonia [41], were also considered. A uniform monthly production of manure was assumed, not affected by the season.

### 3.4 Simulation

To solve the manure transfer problem from livestock farms to crop fields in an optimal way, a centralized optimized approach (COA) has been developed [37], based on an algorithm that generalizes and adapts the well-known Dijkstra's algorithm for finding shortest paths [42], together with the use of origin-destination cost matrices as applied in the travelling salesman problem for choosing best routes [43]. COA solves the problem by considering a shortest-path problem on an undirected, non-negative, weighted graph. COA is described in [37]. To use the algorithm within the context of the problem under study, the algorithm has been modified to respect the necessary configurations and constraints, i.e. by modelling the weights of the graph to represent both transport distances and crop farms' nitrogen needs. All combinations of visits to nearby farms (within a wide radius of 100 kilometers) are added to an origin-destination cost matrix, where the most profitable route is selected. In contrary to the typical travelling salesman problem, here the possible stop locations vary depending on which combinations of candidate crop farms maximize the following global objective (GO):

$$GO = (NT * 0.225 * l) - (TD * 0.1827) \qquad (1)$$

where *NT* is the total nitrogen transferred in kilograms at every *transaction*, and *TD* is the total distance in kilometers covered to transport manure from the livestock farm to the crop field. The parameter *l* captures the nutrient losses of manure during its storage time, i.e. the time when the manure is stored at the livestock farm until it is transferred to the crop field. A loss of 5% has been considered based on [44].

A decentralized, nature-inspired approach based on an ant colony optimization algorithm has been published by the authors as well [38], but with 8% less performance (in terms of the global objective GO) in comparison to COA.

# 4 Results

This section presents the findings obtained by solving the problem of manure transport optimization, based on the modelling and the simulations performed in Section 3.

Table 2 summarizes the total carbon stored at each different crop type, based on the transfers that took place after running the simulator for a complete one-year duration. A total of 106 thousand tons of carbon is indirectly stored into 10,982 fields of 19 different crop types, spread at a total area of 390 thousand hectares. The mean value of carbon added in each crop field where animal manure was applied was 0.29 kg C/m2 as a consequence of applying mean doses of animal fertilizer around 1.34 kg/m2.

**Table 2.** Kilograms of carbon (in thousands) stored in different types of crop fields in the hypothetical scenario of transferring animal manure from livestock farms to crop fields, after a year.

| Crop | Month | | | | | | | | | | | |
|---|---|---|---|---|---|---|---|---|---|---|---|---|
| | 1 | 2 | 3 | 4 | 5 | 6 | 7 | 8 | 9 | 10 | 11 | 12 |
| ALFALFA | 0 | 0 | 783 | 3184 | 1541 | 102 | 239 | 164 | 0 | 0 | 0 | 0 |
| CORN | 1677 | 1449 | 1433 | 209 | 55 | 60 | 39 | 2 | 0 | 0 | 0 | 397 |
| CEREALS | 441 | 318 | 235 | 31 | 0 | 0 | 0 | 5545 | 20443 | 4386 | 4035 | 593 |
| VEGETABLES | 407 | 161 | 345 | 201 | 107 | 113 | 62 | 57 | 40 | 33 | 37 | 35 |
| OLIVE TREES | 1606 | 1397 | 49 | 21 | 22 | 12 | 5 | 0 | 0 | 0 | 1 | 3 |
| SWEET FRUITS | 1271 | 965 | 25 | 14 | 4 | 9 | 0 | 0 | 0 | 0 | 230 | 2971 |
| WOOD-BASED CROPS | 790 | 459 | 516 | 310 | 270 | 301 | 157 | 110 | 76 | 55 | 90 | 108 |
| VINEYARDS | 535 | 465 | 4 | 3 | 2 | 3 | 0 | 0 | 0 | 0 | 667 | 712 |
| NUT TREES | 971 | 713 | 7 | 3 | 1 | 0 | 0 | 0 | 0 | 0 | 585 | 1543 |
| FALLOW | 0 | 0 | 0 | 0 | 0 | 0 | 0 | 0 | 0 | 0 | 0 | 0 |
| FORAGE | 3802 | 2918 | 2752 | 2233 | 1831 | 1996 | 1470 | 1250 | 773 | 725 | 684 | 1024 |
| SUNFLOWER | 383 | 270 | 212 | 17 | 10 | 7 | 4 | 0 | 0 | 0 | 0 | 59 |
| RAPESEED | 234 | 169 | 27 | 0 | 0 | 0 | 3060 | 566 | 182 | 181 | 381 | 101 |
| LEGUMES | 456 | 168 | 186 | 134 | 65 | 94 | 62 | 28 | 12 | 6 | 6 | 16 |
| OTHER CROPS | 681 | 439 | 513 | 403 | 330 | 288 | 204 | 173 | 69 | 76 | 83 | 117 |
| CITRUSES | 162 | 123 | 0 | 0 | 0 | 0 | 0 | 0 | 0 | 0 | 16 | 731 |
| HAZELNUT | 144 | 141 | 7 | 1 | 1 | 1 | 0 | 0 | 0 | 0 | 168 | 191 |
| HEMP | 30 | 15 | 12 | 4 | 6 | 4 | 4 | 4 | 4 | 4 | 4 | 6 |
| RICE | 48 | 47 | 52 | 50 | 53 | 39 | 35 | 32 | 34 | 33 | 29 | 43 |

| | | | | | | | | | | | |
|---|---|---|---|---|---|---|---|---|---|---|---|
| SOYA | 34 | 20 | 19 | 17 | 9 | 5 | 8 | 4 | 6 | 5 | 4 | 2 |
| CAMELINA | 0 | 0 | 0 | 0 | 0 | 0 | 0 | 0 | 0 | 0 | 0 | 0 |

Figure 2 visualizes the results of Table 1 in a heatmap, to highlight the differences in SOC accumulated in different crop types. As Figure 2 shows, cereals have the most accumulation during the months of August-November. Some worth-mentioning SOC storage occurs also between January-March for forage, July for rapeseed, April for alfalfa and December for sweet fruits and olive trees.

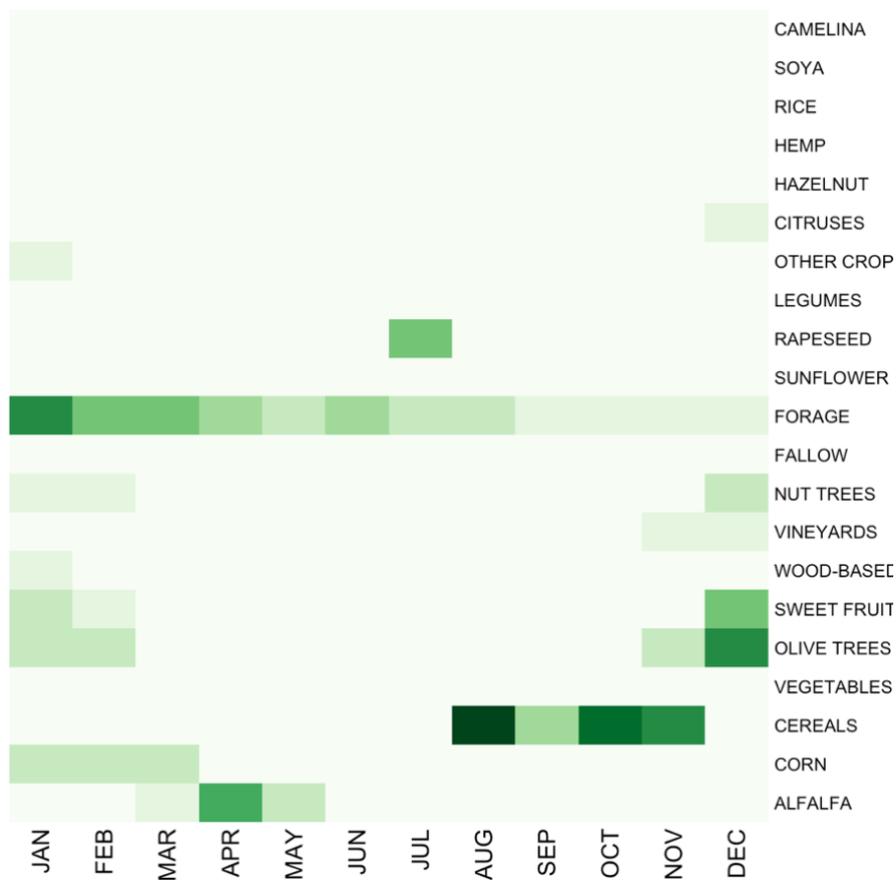

**Fig. 2.** Heatmap visualizing SOC accumulated in different crop types, in different year months.

The map in Figure 3 visualizes the crop fields that satisfy the 4x1000 strategy after the first year of the approach under study is applied. These crop fields are depicted in green color, in comparison to the rest shown in red color. Most of the crop fields satisfying the strategy are concentrated in the regions of Lleida and Girona, where livestock farms are highly concentrated thus the transfer of manure is possible and efficient, due to the short distances between the farms and the fields.

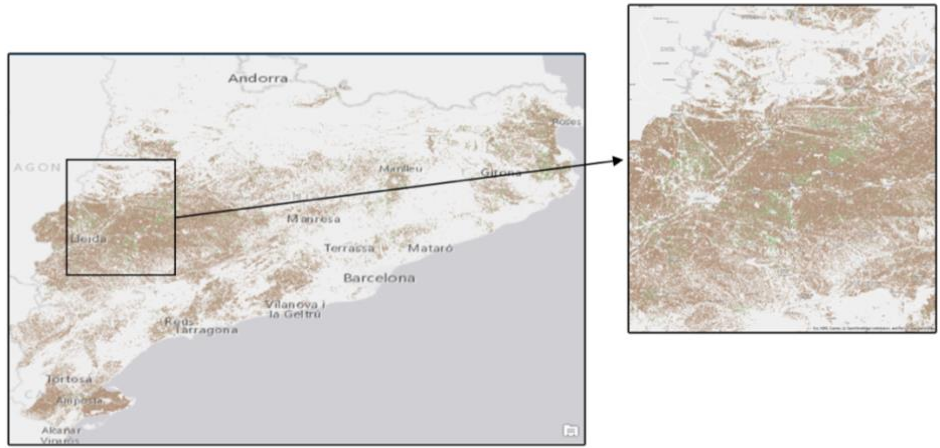

**Fig. 3.** Map of Catalonia showing the crop fields of the region. The fields satisfying the 4x1000 strategy at the first year of the manure transfer program are shown in green color, the rest fields in red color. The region near the city of Lleida is magnified at the right side.

## 5    Discussion

The present study found that applying animal manure as a single strategy to increase SOC stocks in Mediterranean agricultural soils in North eastern of Spain would be not enough to reach the 4x1000 goals in a duration of one year[1]. Therefore, it will be necessary to assess other recommended management practices to sequestrate carbon in agricultural soils. In fact, some studies highlighted the potential of combining practices to improve sequestration [16-18]. Our study has some limitations that are important to be mentioned:
  a) SOM for some animal types has been calculated in several studies outside Europe (India, China), as well as in Catalonia, at different soil depths.
  b) The effect of manure management systems (treatment units for pig slurry, compost units for bovine manure) was not considered.
  c) Only the first year of manure application is actually assessed. Long time periods have been considered only naively, without considering weather/climate modelling/forecasting.

Climate variables and soil properties are important drivers of SOC dynamics [39], but have not been considered in the calculations, even in the first-year case of applying the proposed approach. It is widely known that climate variables are important drivers of SOC stock: increasing SOC is associated with higher annual precipitation and lower temperature [45-46]. Higher temperatures are associated to higher rates of mineralization due to the increase of microbial activity. However, soil moisture could act as the main driver of soil microbiomes in Mediterranean environments, limiting

---

[1] Further analysis not shown in this paper shows that the 4x1000 goal cannot be reached by more than 60% of the farms even after 10 years.

SOC losses by microbial mineralization [47]. Otherwise, a decrease in available soil water content would negatively affect yields and, consequently, the associated soil carbon input. Soil properties can also affect SOC stocks as much as organic carbon is stabilized by means of physical protection or chemical mechanisms [48].

The effects of climate change on global soil carbon stocks are controversial [8]. However, it seems clear from the environmental projections that the Mediterranean is warming at a pace that is 20% faster than the global average, leading to a possible regional increase of 2.2 ºC of temperature and a 15% reduction in rainfall by 2040, considering that current policies would still be in place by then. The combination of these weather events will promote a high evapotranspiration and, consequently, an increase in the duration and frequency of droughts [5].

The obtained results in this study are important for the 4x1000 strategy under the prism of our actual agricultural system, because it indicates that the soil can improve the carbon balance in specific farming regions and scenarios (see Figure 3), but it can only contribute partly. The fact of improving only certain regions could be associated to the concentration of livestock farms in specific areas and to the fact that pig slurry, containing low carbon content (Table 1), is the major animal fertilizer produced in Catalonia. In order to increase the possibilities to achieve the 4x1000 strategy and prevent ground water pollution by nitrogen, it would be necessary to apply composted manure that contains higher content of carbon. It is true that soil is a very important carbon sink, but the application of 4x1000 must be accomplished according to new integrated models of agriculture, in which sources and sinks of carbon will be closed. The method of transferring manure needs to be combined with other strategies, such as those cited in Section 2 (Related work) [14-32]: cover crops, crop rotation, reduced tillage, crop residue incorporation to the soil, among others.

### 5.1 Future work

The results of the present study indicate that merely with the application of manure in agricultural soils, the 4x1000 strategy could not be reached in a reasonable number of years for all the existing crop fields. Hence, future efforts should focus on assessing the upscaling of different management practices that are proved to sequestered carbon in the soil by themselves and the combination of different practices. Moreover, it is important for future work to consider the effects of:

   i.   Precipitation/temperature changes using different climate change scenarios;
   ii.  Land uses changes;
   iii. More realistic models to calculate the SOM based on animal manure;
   iv.  Manure management systems and multiple years; and
   v.   $CO_2$ emissions of manure transportation from livestock farm to crop field.

## 6     Conclusion

Soil organic carbon (SOC) plays an important role in improving soil conditions and soil functions. Increasing land use changes have induced an important decline of SOC content at global scale. Animal manure has the characteristic of increasing SOC, when applied to crop fields, while, in parallel, it constitutes a natural fertilizer for the crops.

In this paper, we have developed a large-scale simulation, using the whole area of the region of Catalonia (Spain) as a case study. The goal was to examine whether animal manure can improve substantially the SOC of the Catalan crop fields. Our results showed that the policy goals of Spain can only be achieved partly by using merely manure transported to the fields, thus this strategy needs to be combined with other agricultural practices. We have discussed implications of the findings, together with additional policies, actions and methods required in order to reach the policy goals of Spain in terms of enriching the SOC in Spanish soils, mitigating the climate change effects.

## Acknowledgments

This research was supported by the CERCA Programme/Generalitat de Catalunya. The assistance of the Catalan Ministry of Agriculture, Livestock, Fisheries and Food is also acknowledged. Francesc Prenafeta-Boldú belongs to the Consolidated Research Group TERRA (ref. 2017 SGR 1290).

Andreas Kamilaris has received funding from the European Union's Horizon 2020 research and innovation programme under grant agreement No 739578 complemented by the Government of the Republic of Cyprus through the Directorate General for European Programmes, Coordination and Development.